\documentclass[%
rsi,
 amsmath,amssymb,
reprint,%
]{revtex4-1}

\usepackage{graphicx}
\usepackage{dcolumn}
\usepackage{bm}
\usepackage{xcolor}
\usepackage[utf8]{inputenc}
\usepackage[T1]{fontenc}
\usepackage{mathptmx}

\begin{document}

\preprint{AIP/123-QED}

\title[]{AC Measurement of the Nernst effect of thin films at low temperatures }

\author{Y. Wu}
\address{Department of Physics, Jack and Pearl Resnick Institute the and Institute of Nanotechnology and Advanced Materials, Bar-Ilan University, Ramat-Gan 52900, Israel}
\author{S. Dutta}
\address{Tata Institute of Fundamental Research, Homi Bhabha Road, Mumbai 400005, India}
\author{J. Jesudasan}
\address{Tata Institute of Fundamental Research, Homi Bhabha Road, Mumbai 400005, India}
\author{A. Frydman}
\address{Department of Physics, Jack and Pearl Resnick Institute the and Institute of Nanotechnology and Advanced Materials, Bar-Ilan University, Ramat-Gan 52900, Israel}
\author{A. Roy}
\email{res.arnab.roy@gmail.com}
\address{Department of Physics, Jack and Pearl Resnick Institute the and Institute of Nanotechnology and Advanced Materials, Bar-Ilan University, Ramat-Gan 52900, Israel}

\date{\today}

\begin{abstract}

We describe an alternating current method to measure the Nernst effect in superconducting thin films at low temperatures. The Nernst effect is an important tool in the understanding superconducting fluctuations and, in particular, vortex motion near critical points. However, in most materials, the Nernst signal in a typical experimental setup rarely exceeds a few $\mu$V, in some cases being as low as a few nV. DC measurements of such small signals require extensive signal processing and protection against stray pickups and offsets, limiting the sensitivity of such measurements to $>$1nV. Here we describe a method utilizing a one-heater-two-thermometer setup  with the heating element and thermometers fabricated on-chip with the sample, which helped to reduce thermal load and temperature lag between the substrate and thermometer. Using AC heating power and 2$\omega$ measurement, we are able to achieve sub-nanovolt sensitivity in 20-30nm thin superconducting films on glass substrate, compared to a sensitivity of $\sim$10nV using DC techniques on the same setup.
\end{abstract}

\maketitle

\section{\label{Introduction}Introduction}
Over the past few decades, thermoelectric effects have emerged as powerful tools to study the electronic properties of materials. Being intimately connected to the electrical conductivity $\sigma$ and thermal conductivity $\kappa$, the thermoelectric conductivity $\alpha$ provides different, but complimentary information. In the case of metals and semiconductors, the diagonal component $\alpha_{xx}$ of the thermoelectric tensor $\tilde{\alpha}$ is the dominant thermoelectric response \cite{behnia2015fundamentals,Behnia2016}. The off diagonal elements, which typically appear only in the presence of a magnetic field, are usually small due to generic symmetries of the Fermi surface. In the case of superconductors, the dominant response is the off-diagonal component $\alpha_{xy}$, and the diagonal response is vanishingly small. This is because in superconductors, the source of the thermoelectric effects is primarily superconducting fluctuations rather than quasiparticles as in the case of metals. Since superconducting fluctuations are carriers of entropy, they travel down the temperature gradient, generating a transverse phase-slip voltage in the process. Being a signal arising purely from superconducting fluctuations, measurement of $\alpha _{xy}$ becomes particularly interesting in the field of 2D disordered superconductors. 

The physical property that is accessible in a thermoelectric measurement is the Nernst effect, i.e. the transverse voltage per unit temperature gradient in the presence of perpendicular magnetic field: $ N = \frac{\partial V _y}{\partial \nabla T} \big| _{\nabla T || x ; B || z}$ and the Nernst coefficient $\nu=\frac{\partial N}{\partial B}\big| _{B \rightarrow 0}$. In the case of superconductors which are characterized by an effective particle-hole symmetry, the Nernst coefficient can be expressed as a product of the resistivity and the transverse Peltier coefficient $\nu=\rho_{xx} \cdot \alpha_{xy}$. During the past few years the Nernst effect has been shown to be a very effective tool to study vortex motion in the fluctuation regime both above and below $T_C$ \cite{Wang2002,Wang2006,Pourret2006,Pourret2007,Spathis2008,Roy2018}. 

In 2D superconductors Nernst effect measurements become especially appealing due to several exotic vortex phases which are expected in these thin films. These system often undergo a direct superconductor to insulator transition as a function of thickness or disorder. Experiments have revealed signs for cooper pairing  and a finite energy gap in the insulating phase. Such a phase has been dubbed a “Bosonic insulator”. An example is a thin film of amorphous indium oxide a-InO${_x}$ in which evidence for vortices motion or a finite energy gap have been detected in the insulator \cite{Poran2011,Kopnov2012,Sacepe2011,Sherman2012,Sherman2013}. Indeed, significant Nernst coefficients have been measured in both the superconducting and the insulating phases \cite{Roy2018}.  

Experiments on other systems \cite{Aubin2006,Marrache-Kikuchi2008,Tsen2016,Kapitulnik2019} and theories \cite{Das1999,Dalidovich2001} have also raised the possibility of an intermediate anomalous “Boson metal”  phase between the insulator and the superconductor. Such a phase, that contradicts the accepted notion that a 2D metallic state cannot exist, is under heavy deliberations nowadays \cite{Tamir_2019} and Nernst measurements may assist in the elucidation of this issue.  

In yet other systems like MoGe, an exotic hexatic vortex fluid is encountered on the approach to the zero-resistance state in the presence of a magnetic field \cite{Roy2019}. By studying the field dependence of the Nernst signal in systems like these, the characteristic length scales associated with superconductivity can be extracted, giving information about the underlying physcial processes both at the microscopic and mesoscopic level. 

In high-T$_c$ cuprates, the Nernst effect has been used extensively to attempt to understand the nature of the pseudogap state \cite{Wang2002,Wang2006}. The nature of the pseudogap state of High-T$_c$ cuprates like YBCO is being hotly debated even to this day, with suggested explanations diverging between two key ideas, a fluctuation-dominated pseudogap phase with preformed Cooper pairs, and a competing-order hypothesis that describes the pseudogap state as a competing ground state with a magnetic order. Nernst effect, originating purely out of superconducting fluctuations has played a key role in being complimentary to electrical conductivity measurements, and has helped to refine the phase diagrams of many such systems \cite{Wang2002,Ri1994,Tafti2014}.

In semimetals like bismuth and graphite, Nernst effect is of electronic origin, and shows Landau quantization in the form of oscillatory response to a magnetic field in the lines of de Haas-van Alpen and Shubnikov-de Haas oscillations \cite{Behnia2007,Mangez1976,STEELE1955,Zhu2010,Zhu2011}. Based on these observations, Nernst effect has emerged as a new tool to study the Fermi surface properties of 2d materials like graphite \cite{Checkelsky2009}. Anomalous Nernst effect, like anomalous Hall effect may arise even at zero magnetic field in materials such as Weyl semimetals, which have non-zero Berry curvature of the reciprocal space \cite{Xiao2006,Watzman2018,Sakai2018,Guin2019,PhysRevLett.118.136601}. A special form of the anomalous Nernst effect is observed in materials with strong spin-orbit interaction. Termed as the spin-Nernst effect \cite{Meyer2017,Sheng2017}, it is an accumulation of electronic spins, instead of charge in a direction transverse to the heat flow, caused by the differential scattering of up-spin and down-spin electrons by the heavy element atoms.

Thus clearly, the Nernst effect has a rich phenomenology and is an important tool to study the physics of several classes of materials. It should be noted that the Nernst response $N$ typically varies over many orders of magnitude with temperature but seldom exceeds a few $\mu$V/K. Information is often contained in regimes where the signal is only a few nV, which makes precise measurement of utmost importance. 

Two classes of Nernst effect measurement are encountered in literature depending on whether lockin techniques are used for the voltage measurement. DC measurements, involving a constant heating current and constant temperature gradient are realtively simple to interpret, but come with the additional complexity of compensating for stray thermoelectric voltages as well as extreme sensitivity to environmental and instrumental noise.  Noise levels are kept within acceptable levels by careful shielding and filtering and require the use of special low-noise electronics. Stray thermoelectric voltages originating in the measurement leads are minimized by avoiding junctions altogether in the measurement path. Using this method, a signal detection threshold of 5 nV \cite{Wang2006} or even 1 nV \cite{Pourret2006} has been achieved. 

Following the original suggestion by Corbino \cite{Corbino1910} of using higher harmonics of the alternating current in an incandescent lamp to study properties of the filament, the use of higher harmonics like $2\omega$ or $3\omega$ has now emerged as a powerful technique to study material properties that can be modulated through temperature oscillations. Some of the most sensitive measurements are those involving specific heat ($2\omega$ or $3\omega$) \cite{Lu2001,Poran2014}, thermal conductivity ($3\omega$) \cite{Lu2001,Sikora2012,Mishra2015} and thermoelectric effects ($2\omega$) \cite{Kettler1986,Oussena1992,Choi2001}. AC measurements using lockin techniques are somewhat less demanding with regard to the requirement of noise shielding and special low-noise amplifiers, but their interpretation is complex and usually require elaborate calibration procedures. In the context of the Nernst effect, the use of low-frequency alternating currents for heating and detecting the $2\omega$ signal by lockin techniques, has resulted in a signal detection threshold of 0.5nV \cite{Kettler1986}.

Thus, several factors need to be considered to make accurate measurements of the Nernst effect. (i) The sample and measurement leads should be shielded from ambient electromagnetic radiation (ii) Thermoelectric voltages at junctions of measurement wires must be carefully compensated (iii) A large-enough temperature gradient should be generated for producing measurable signal (iv) The temperatures and temperature gradients must be accurately determined. This is achieved by making heat transfer between the substrate and the heater/sample/thermometers efficient. In the present article we describe a method suitable for the measurement of magneto-thermoelectric effects of thin film samples at low temperatures in the range $\sim$ 0.3K to 10K. The novelty of our technique rests in the fact that ac techniques have not been used so far in the low-temperature regime, where the time scales for heat transfer increase dramatically due to dropping thermal conductivities. Because slow heat transfer rates are a central problem in the use of ac techniques at low temperature, we use lithographically fabricated on-board thermometers instead of commercial calibrated temperature sensors. This has enabled us to use a relatively high frequency of heating (2Hz) without generating significant thermal lags, which has helped significantly in achieving a remarkable signal detection threshold of $\sim$1 nV despite having a DC noise level as high as 30nV in the same setup.

\section{\label{Experiment}Experiment}

\subsection{Samples}

\begin{figure}[h]
\vspace{0cm}
\centering
\includegraphics[width=0.5\textwidth]{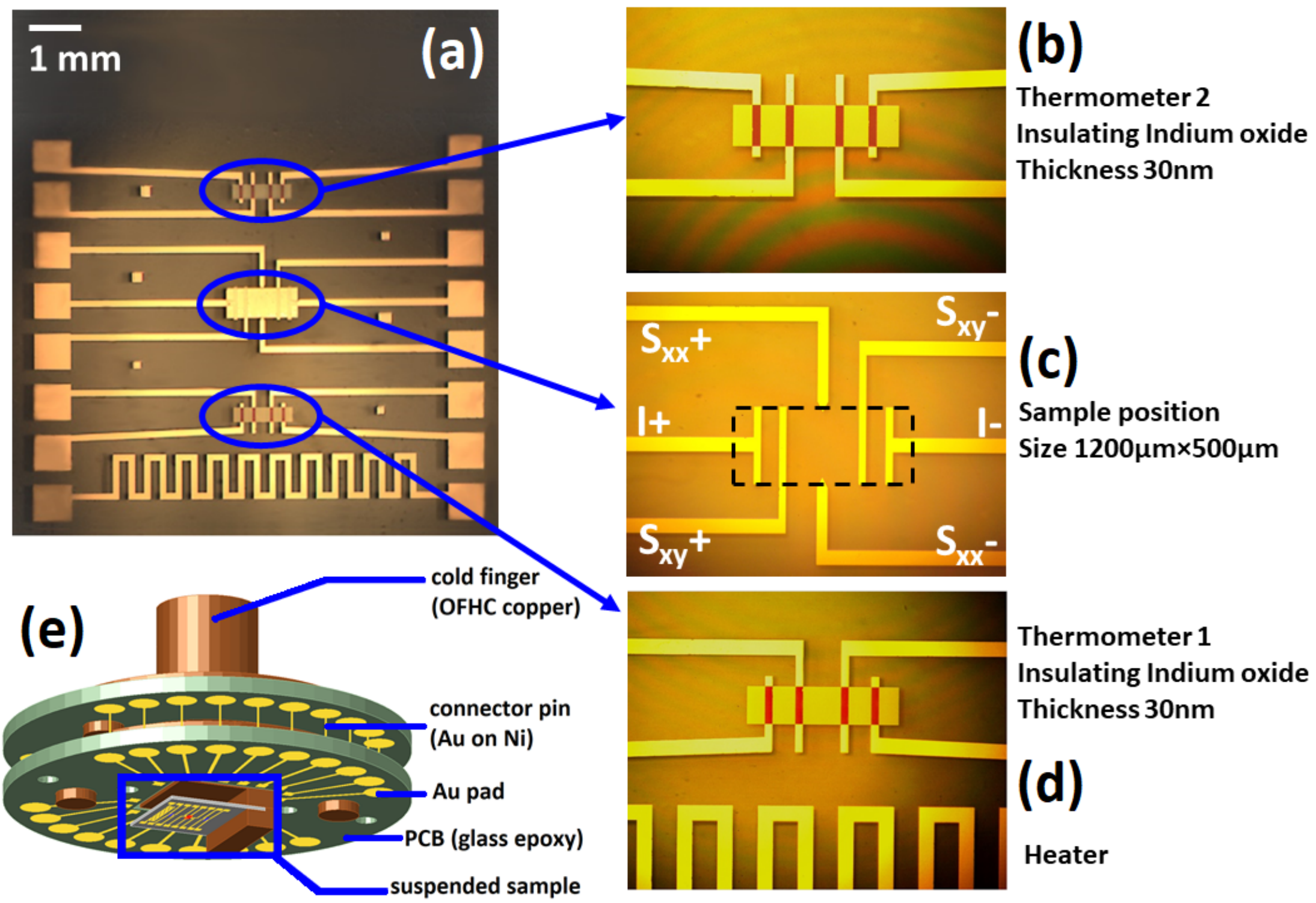}
\caption{(color online) (a) Optical image of the chip with 4 devices: 2 thermometers, 1 heater and 1 sample. (b) and (d) Magnified images of the two thermometers, 30nm thick insulating Indium oxide with leads for four-terminal resistance measurement. (c) Magnified image of the sample position, with leads for measurement of Seebeck ($S_{xx}+$,$S_{xx}-$) and Nernst ($S_{xy}+$,$S_{xy}-$) leads. Seebeck and Nernst leads double as Hall and resistance leads respectively when current is passed between $I+$ and $I-$. (e) Chip carrier  on cold finger with the Nernst chip.}.
\label{chip}
\end{figure}

The thermoelectric setup, comprising a heater, two thermometers and the sample are fabricated on a chip of MEMpax\texttrademark  borosilicate glass substrates of dimension 1cm$\times$1cm$\times$0.3mm by optical lithography, (Fig. \ref{chip}). A gold meander serves as a heater. It is fabricated out of 30nm thick Au with a 4nm underlayer of Cr, and is designed to have a resistance of $200 \Omega$ at room temperature. The two thermometers are fabricated from e-beam evaporated Indium Oxide, with the growth parameters being tuned to achieve moderately insulating behaviour ($R _{\square} \sim 10-200k \Omega  $ at low temperature). The sample is grown by standard thin film deposition techniques like thermal/ebeam evaporation, pulsed laser ablation etc. Layered 2d materials can also be used as samples using the van der Waals transfer method with suitable modifications to the electrical leads. The choice of substrate is dictated by the requirement of having a very low thermal conductivity which enables the setting up of a large enough heat current without the application of excessive heating power. Glass has a thermal conductivity in the range of $ 0.01 W m^{-1} K ^{-1}$ at 1K, among the lowest of all materials. Glass is also a natural choice when the films to be grown are of amorphous nature. The glass substrate is suspended from the cold finger in the manner shown in Fig.\ref{chip}e, thermal contact being provided on the substrate edge far from the heater. 

\begin{figure}[ht]
\vspace{1cm}
\centering
\includegraphics[width=0.45\textwidth]{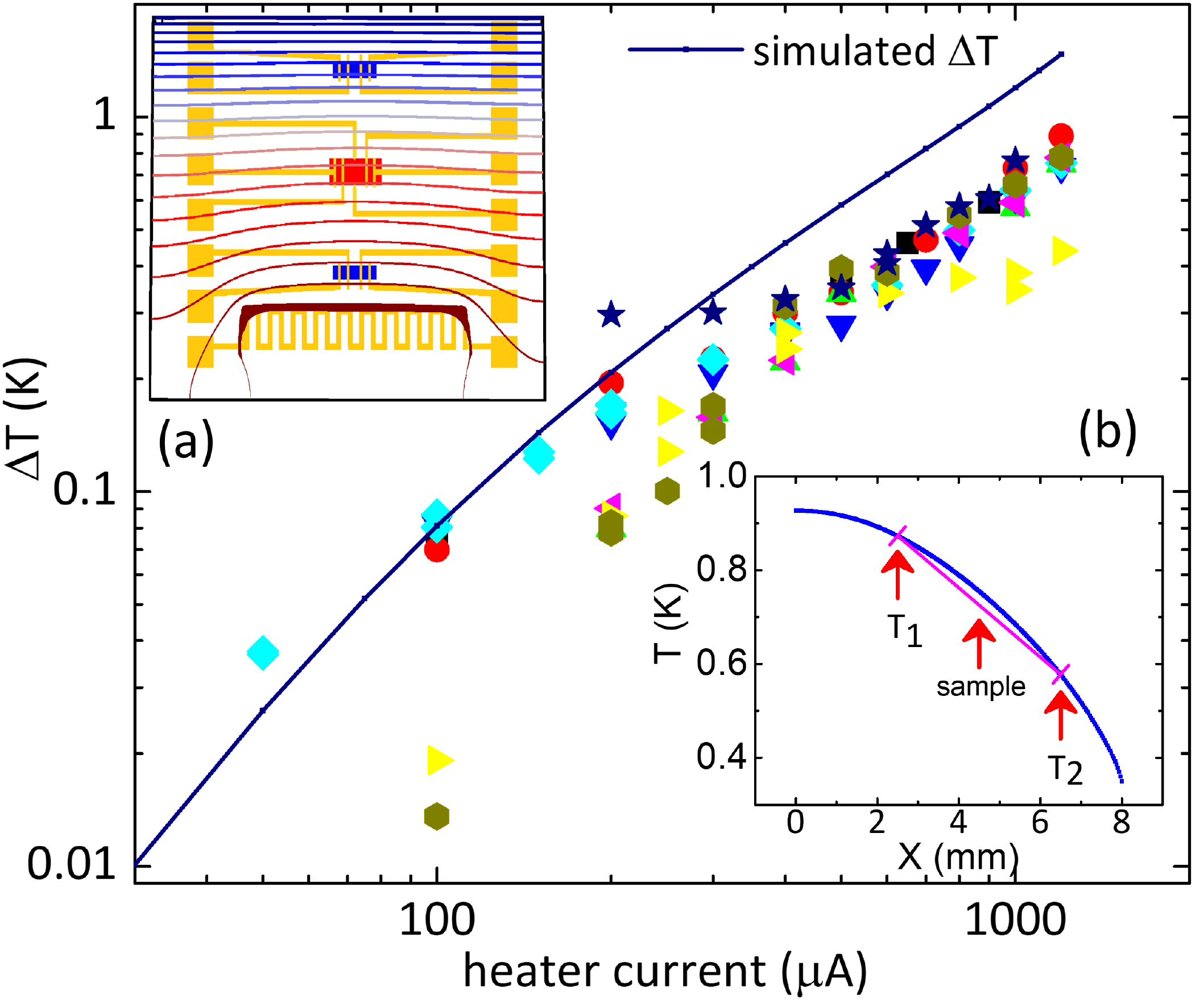}
\caption{(color online) Temperature difference between two thermometers as a function of heater current (DC) : Comparison between experimental points (symbols, each color indicating a different sample) and finite element modelling (blue line). Inset (a) Isothermal contours with a 1 mA heater current obtained by finite element modelling. Inset (b) Simulated temperature profile in a direction parallel to the heat current. The straight line indicates the region of linear approximation used for experiment.}
\label{dT}
\end{figure}

\subsection{Temperature measurements}

In the absence of standard calibration curves, the thermometers are calibrated through a measurement of R-T curves against calibrated sensors in the cryostat prior to the measurement. Both AC and DC measurement techniques are employed. For DC measurements, two pairs of Yokogawa 7651 current sources and Keithley 2000 multimeters are used. The current is kept in the range of 10 - 100nA. For AC measurements, two SR830 lockin amplifiers are used, the current being 10nA. Temperature gradients are then measured as a function of the applied heater current, which forms the second part of the calibration in the case of DC measurements. $\Delta T _{DC} = T ^{1} _{DC} - T ^{2} _{DC}$, $\nabla T _{DC} = \Delta T _{DC}/ (X_2 -X_1)$. In Fig.\ref{dT}, the temperature difference between the two thermometers is plotted as a function of current through the heater for different samples prepared identically along with a simulated curve generated from a finite element simulation. For the simulation COMSOL finite element package was used \cite{Note2}, with tetrahedral elements having a maximum mesh size of 240$\mu m$. Adiabatic boundary conditions were chosen for the free surfaces of the sample and isothermal conditions for the interface with the cold finger. Low-temperature thermal conductivity and specific heat were modelled using values obtained from \cite{R.B.Stephens1973,Note1}. Calculated isotherms indicated no transverse temperature gradients at the sample position (Fig.\ref{dT}a) and very small non-linearity in the temperature gradient between the two thermometers at X = 2.5 mm and X = 6.5 mm (Fig.\ref{dT}b). 

\begin{figure}[ht]
\vspace{1cm}
\centering
\includegraphics[width=0.45\textwidth]{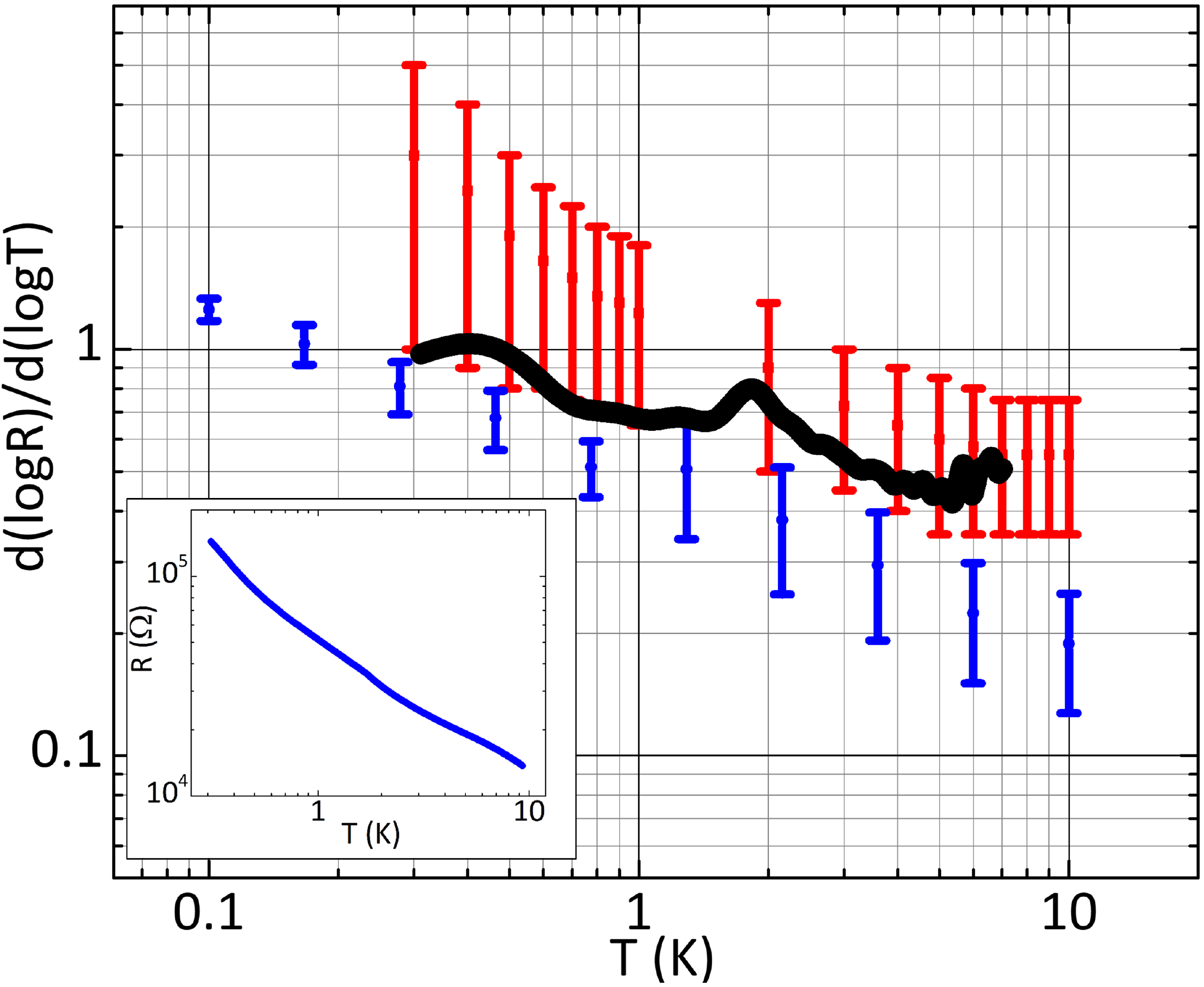}
\caption{(color online) Sensitivity $ \frac{T}{R} \frac{dR}{dT}$ of a typical Indium Oxide thermometer (black line) compared with two common types of thermometers that are used in this temperature range: Cernox\texttrademark and Ruthenium oxide. Inset: R vs T curve of the same thermometer. Red bars indicate the range of sensitivities of Cernox sensor CX1030. Blue bars indicate the range of sensitivities covered by 3 types of RuO\textsubscript{2} sensors, RX-102A, RX-103A and RX-202A. Data was reproduced from Temperature Measurement and Control Catalog, Lake Shore Cryotronics, Inc. (2016). }
\label{sensitivity}
\end{figure}

Like in any experiment, the measurement of temperatures and temperature gradients are subject to two types of errors, random and systematic. Random errors in the form of measurement noise can be estimated from the sensitivity $\lvert\frac{T}{R} \frac{dR}{dT}\rvert$ of the sensor at various temperatures (Fig.\ref{sensitivity}). Our resistance measurement noise $\frac{dR}{R}\vert _{noise}$ being a constant $5 \times 10^{-3}$ between 0.5K to 10K enabled us to estimate a temperature readout error of $\sim 2.5 mK$ at $0.3K$ and $\sim 25 mK$ at $3K$.  By adopting AC technique of measurement of thermometer resistances, the noise level $\frac{dR}{R}\vert _{noise}$ can be reduced to $3 \times 10^{-4}$, improving sensitivity by an order of magnitude to $0.2 mK$ at $0.3K$. However its relevance to temperature measurement accuracy is unclear owing to a lack of detailed knowledge on the reproducibility of thermometer characteristics.  It may be noted here that commercial Cernox\texttrademark sensors are limited to a temperature accuracy of $3-6 mK$ in this temperature range, irrespective of the quality of measurement \cite{Note1}. We are, however, aware of temperature difference measurement accuracy of $0.1mK$ \cite{Pourret2006} and temperature measurement accuracy $\sim 5\mu K $ \cite{Ptak2005} in this temperature range. 

Systematic errors are mainly due to the use of the linear approximation described above. Such errors were estimated from the finite element simulation. At a DC heating current of 1mA, the following numbers were obtained: $T _{DC} (exact) =0.756K; \ T _{DC} (lin.approx.) = 0.726K; 
\ \nabla T _{DC}(exact) =0.0726K/mm; \ \nabla T _{DC}(lin.approx.) =0.0739K/mm$, indicating that for practical purposes, the linear approximation was adequate since the systematic error in the estimate of $\nabla T _{DC}$ was less than $2 \%$. On the other hand, measurement noise adds another $\sim 2 \%$ uncertainty to the temperature gradient, making the total uncertainty $\sim 4 \%$.

\subsection{DC measurement}
\begin{figure*}
\vspace{1cm}
\centering
\includegraphics[width=0.95\textwidth]{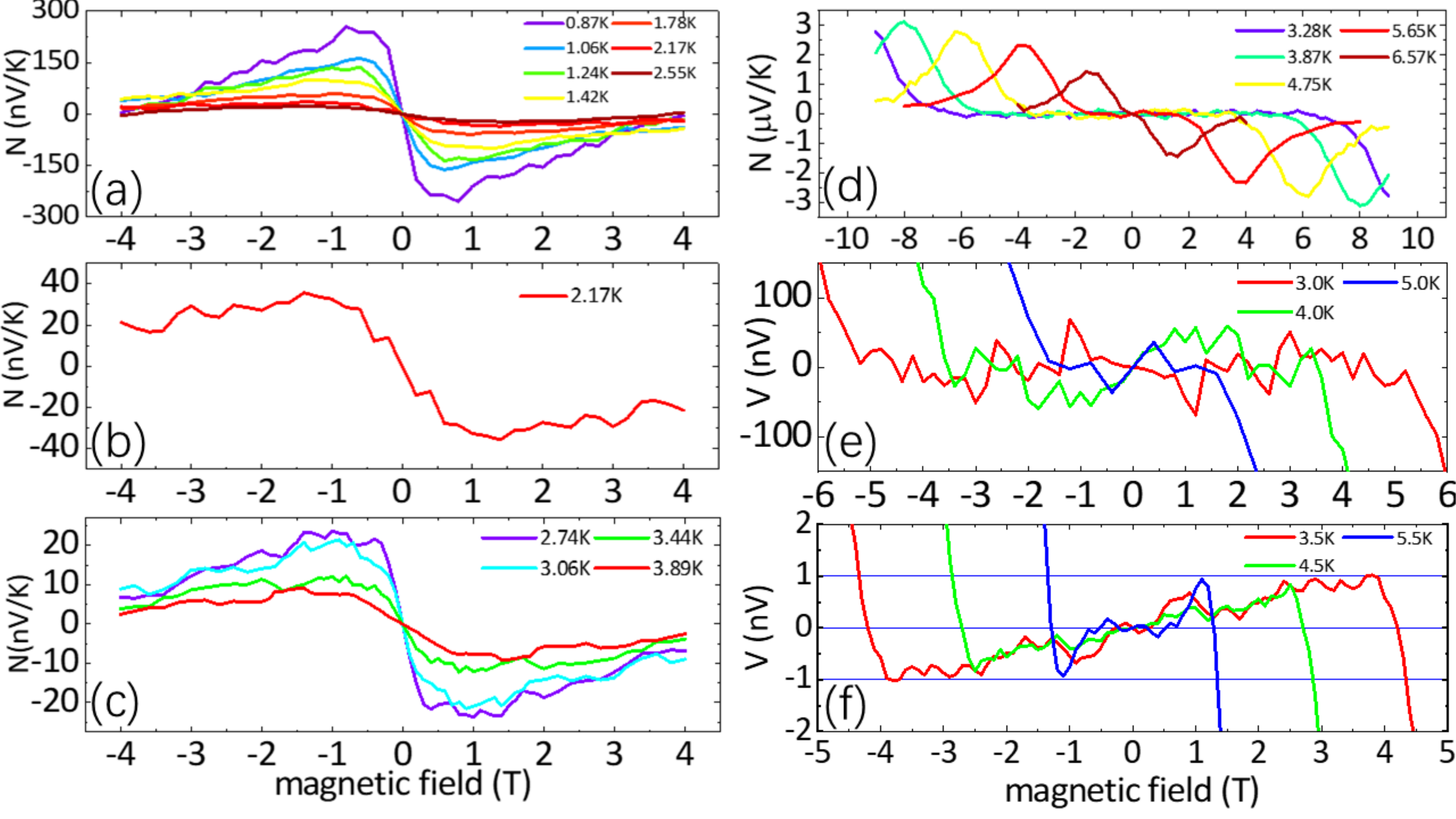}
\caption{(color online) DC and AC Nernst measurements of a 30nm thick weakly disordered superconducting a-InOx film(a-c) and a 25nm thick superconducting amorphous MoGe film (d-f). (a) Large scale DC measurements of an a-InOx film.(b) A DC Nernst response near $T_c$ taken from (a). (c) AC Nernst response at higher temperatures. (d) Large scale DC measurements of an anamorphous MoGe film .(e) Zoom into the low field regime of (d). (f) AC Nernst measurement of the low field regime.}
\label{DCAC}
\end{figure*}

To minimize unwanted thermoelectric voltages at metal junctions, two pairs of manganin wires are run directly from the measuring instrument to the chip carrier. The thermoelectric signals are measured via Keithley 182 or Keithley 2182 nanovoltmeters using a relaxation+acquisition protocol. Magnetic field is applied perpendicular to the sample plane with superconduting magnets in a \textsuperscript{3}He cryostat. Since an ordinary Nernst effect is antisymmetric w.r.t. the applied magnetic field, any residual background voltage is removed by numerically anti-symmetrizing the raw data. Fig.\ref{DCAC}a and \ref{DCAC}b show typical Nernst responses of an a-InOx 2D disordered superconductor as a function of temperature and magnetic field.  In this case the Nernst signal is generated by Gaussian fluctuations of the superconducting order parameter close to a disorder-tuned  superconductor insulator quantum phase transition. The asymmetric peaked structure is typical of superconductors in this regime \cite{behnia2015fundamentals}. The rms noise level in these DC measurements is in the range of $\approx 10 nV$. We note that this noise level is higher than the instrumental noise floor. Our samples are highly disordered superconductors on the verge of being insulators which possess intrinsic sample noise due quantum superconducting phase fluctuations, two level systems etc.

Figure \ref{DCAC}d shows the DC measurement of Nernst effect in a 25nm thick MoGe film. Here the signal is generated by the motion of superconducting vortices after the vortex-lattice undergoes a melting transition due to the application of magnetic field \cite{Roy2019}. A zoom into the low field regime (Fig. \ref{DCAC}e) indicates that a finite Nernst effect may be present. However it is overshadowed by the  measurement noise. Hence in this regime the DC measurement technique, which may be adequate for most purposes, is not sufficient  and a more sensitive AC technique is required.

\begin{figure}
\vspace{1cm}
\centering
\includegraphics[width=0.45\textwidth]{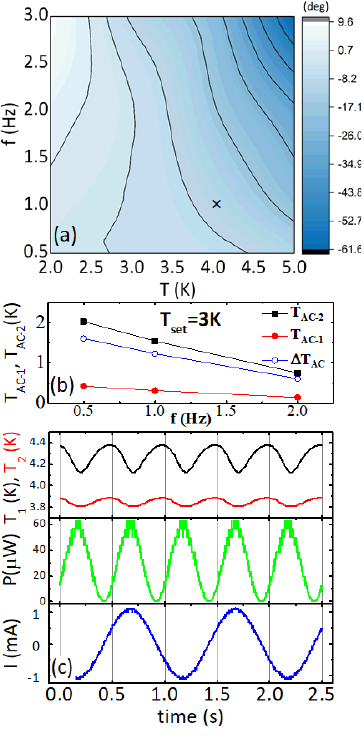}
\caption{(color online)(a) Contours of constant phase difference between two thermometers as a function of heater frequency and cryostat temperature.\textquoteleft$\times$\textquoteright marks the parameter pair where (c) was obtained. (b) AC temperature amplitudes of two thermometers showing that the temperature difference decrease with an increasing heating frequency. Heating current: 5mA (c) Simultaneous measurement of heater current, power and temperature from two thermometers. The plot is obtained from a measurement of the demodulated outputs of two lockin amplifiers measuring thermometer resistances (operating at a much higher modulating frequency) along with the heater driving voltage in an oscilloscope. Though the phase difference between thermometers is small, large phase difference is seen to exist between the calculated heater power and the two temperatures. }
\label{phase}
\end{figure}

\subsection{AC measurement}
This technique involves driving the heater with a sinusoidal current. An AC heating current at frequency $\omega$ generates a heating power which has 2 components: a constant background and an oscillating component with a frequency $2\omega$. As a result the temperature gradients can be divided into two components. The measurement involves detection of the transverse voltage at a frequency of $2\omega$ by lock-in techniques \cite{Kettler1986,Oussena1992,Choi2001}. Calibration of the set-up involves the additional step of measuring the AC component of the temperature gradient $\nabla T _{AC}$. This is done by passing a constant (dc) bias current through the thermometers and recording the voltage amplitude at $2\omega$ with lock-in amplifiers. The lock-in method eliminates higher harmonics ($ >2 ^{nd}$) of the heater current, which arise due to the strong temperature dependences of the thermal conductivity and specific heat of the substrate. When the temperatures of the two thermometers are in phase, $\nabla T _{AC}$ can be simply estimated from the difference in the temperature amplitudes at the two thermometers $\Delta T _{AC} = T ^{1} _{AC} - T ^{2} _{AC}$, $\nabla T _{AC} = \Delta T _{AC}/ (X_2 -X_1)$. However, factors like thermal conductivity and specific heat of the substrate generally cause the phase difference to be non-zero even at low frequencies. So an additional goal of the calibration is to find an optimal frequency that is high enough for a lock-in measurement to be practical, at the same time low enough for phase difference between thermometers to be small. $\omega$ also needs to be low enough for the amplitudes  $T ^{1} _{AC}$ and $ T ^{2} _{AC}$ to be large. This has to be found experimentally, because even the simplest analytical treatment of such a situation gives rise to complicated expressions for the gradient and phase, whose applicability to real-world systems is unclear \cite{Sullivan1968}. The result of one such optimization is shown in Fig.\ref{phase}a. At very low frequencies like 0.5 Hz, the phase difference is small, and increases monotonically with increasing frequency. It is also found to increase with increasing temperature, which probably reflects the different temperature dependences of the two factors, specific heat and thermal conductivity, which determine the relaxation time. $\tau = C _p /K$ \cite{Sullivan1968}.  We choose 1Hz for our measurements, which provides the right compromise between ease of measurement and small phase difference. In Fig.\ref{phase}c, the instantaneous temperatures are plotted together with the heater current and power at 1Hz driving frequency and a setpoint of 4K. Interestingly, even though the phase difference between the two thermometers is minimal, a substantial frequency dependent phase shift exists between the heater power and the two thermometers. This is another effect of the finite thermal relaxation time of the substrate, and causes the Nernst signal to appear almost entirely in the Y-channel of the lock-in amplifier for these settings. 

Measurement of the Nernst signal is carried out using an EG$\&$G 7265 lock-in amplifier and SR552 and SR560 preamplifiers. The former, with an input impedance of $100k \Omega$ and an input noise level of $\sim 4nV / \sqrt{Hz}$ at 2Hz was used as the first stage and is connected directly to the sample leads. Its output is fed to the SR 560 where it is sent through a built-in band-pass filter with a passband of $0.3Hz$ to $10Hz$ and amplified by a factor of about 1000 before being sampled by the lock-in amplifier. The lock-in operation is carried out with a time constant of 10sec, giving it a bandwidth of $\sim 0.02Hz$. Due to the phase shift between the heating power and the temperature gradient as mentioned above, both the X and Y components are recorded. Post measurement, which involves a set of magnetic field levels at a constant temperature and heating current amplitude, the phase of the signal is adjusted to make one of the components as close to zero as possible. The resultant noise level after antisymmetrization is in the range of 1nV.  Fig. \ref{DCAC}f shows the results of the AC Nernst measurement for the low field regime of the MoGe sample. A clear reproducible signal is visible in a region that was completely overshadowed by noise in the DC measurement. With minimal signal processing, a noise level in the range of 300pV was obtained, a vast improvement over the >30nV noise level in the DC measurement for this sample (Fig. \ref{DCAC}e) or the value of 10 nV for an $InO _x$ sample (Fig. \ref{DCAC}a). This is to be compared with the best DC (1 nV) and AC (500 pV) measurements of Nernst effect found in the literature \cite{Pourret2006,Kettler1986}.

\subsection{Artefacts}
One drawback of a one-heater-two-thermometer thermoelectric setup is that it is difficult to control the temperature and temperature gradient independently. For DC measurements, the temperature of the sample is obtained by interpolation between the two thermometers. The same is applicable to an AC measurement when the phase difference between the thermometers is low enough. But there is no easy way of obtaining the $T_{DC}$ and $\nabla T _{AC}$ when higher frequencies are used apart from direct measurement of the demodulated signal with an oscilloscope as shown in Fig.\ref{phase}c. This can make measurements slow at higher frequencies above 2Hz, involving regression methods to estimate the phase difference at $2\omega$. This condition is generally avoided.

\begin{figure}
\vspace{1cm}
\centering
\includegraphics[width=0.45\textwidth]{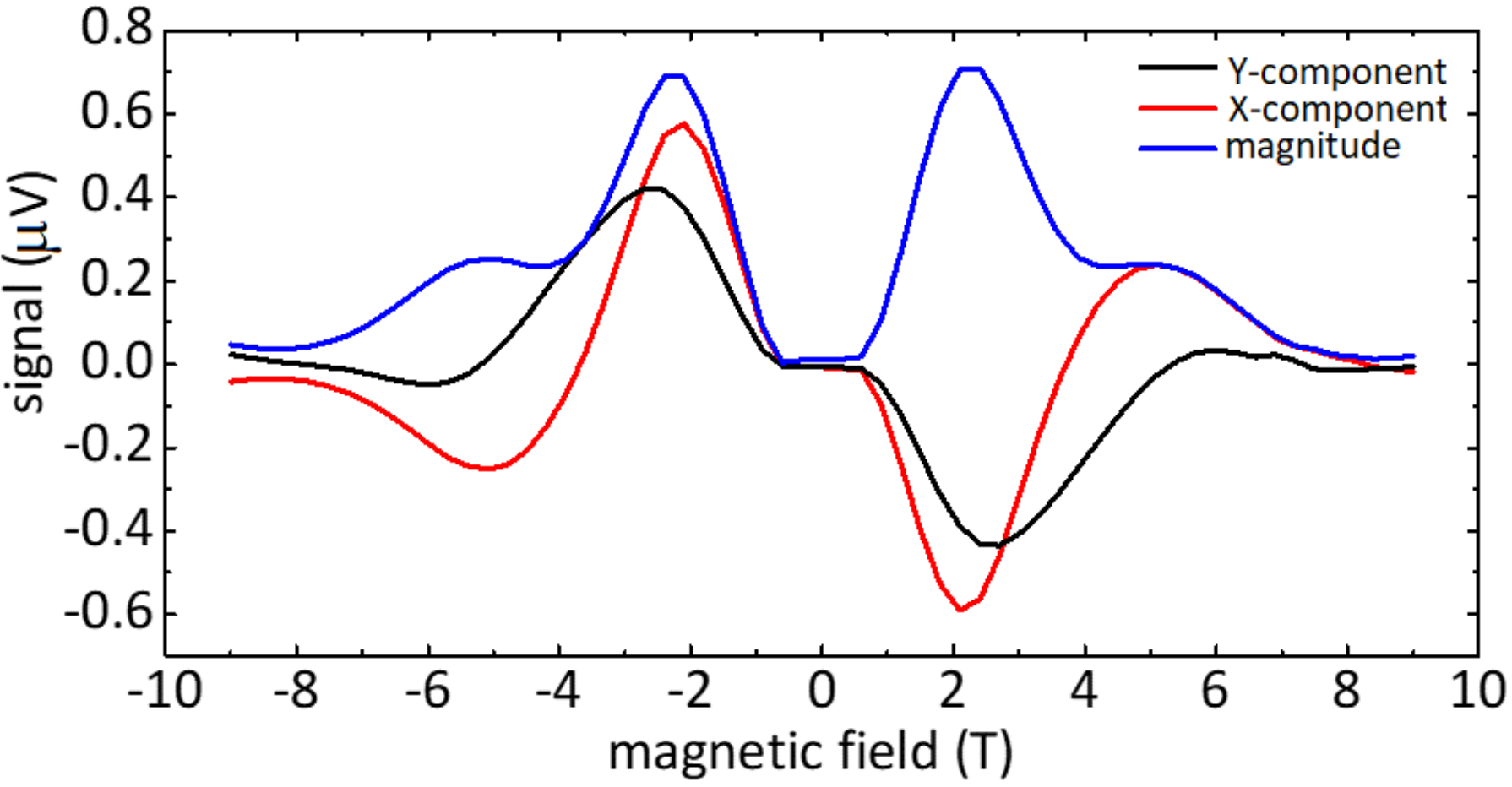}
\caption{(color online) Corrupted Nernst signal as a function of magnetic field due to artefacts mentioned in the text. Sample is MoGe of thickness 25 nm. Heater current is 5mA with a cryostat temperature of 4K.}
\label{artefact}
\end{figure}

A second artefact arises when a current of a large amplitude >$\sim$3mA is used for driving the heater. While this creates a larger $\nabla T _{AC}$ which improves the S/N ratio, a contribution from $T _{AC}$ starts to influence the observed readings. Put simply, the AC voltage measured in a lock-in measurement is composed of two parts: $\frac {dV}{dt}(T,\nabla T) =  \frac{\partial V}{\partial \nabla T} \big| _T \frac{d\nabla T}{dt} + \frac{\partial V}{\partial T} \big| _{\nabla T} \frac{dT}{dt} $. The first part constitutes the Nernst response of the sample, while the second is an artefact with no physical significance. It is small under normal circumstances, but when measuring the Nernst response of a superconductor, the second term can become very important close to the sample's $T_c$ when the resistance of the sample changes rapidly with temperature. There is no easy way of separating these two components, and at high temperatures and magnetic fields, the spurious voltage can dominate the measured transverse voltage. This is illustrated in Fig.\ref{artefact} for an AC heater current of 5mA at a setpoint of 4K, the sample being superconducting MoGe with a $T_c$ of 7K. It can be shown with a simple mathematical treatment that this artefact appears as a voltage orthogonal in phase to the Nernst signal, and is proportional to $\nabla \phi \times T _{AC}$, the product of the phase gradient and the temperature oscillation amplitude. Thus the phase gradient, though small, can cause significant interference to the Nernst measurement at large heating currents, which result in large $T _{AC}$.

\section{Conclusions}

We have developed a thermoelectric measurement setup suitable for thin films at low temperatures. With the heater and thermometers fabricated on-chip, the setup minimizes thermal lags, thus enabling accurate measurements of temperatures and reliable AC measurements. We have developed comprehensive protocols for measurements in DC and AC modes. In AC mode with severely restricted bandwidth, a noise level of 0.3-1 nV is obtainable, much smaller than the noise level of the DC method applied on the same sample.

An array of technical considerations and artefacts limit the useful regime to $0.5 Hz$ to $2Hz$ for glass substrates below 10K. Though with our present measurement apparatus it is possible to achieve 1nV resolution regularly, there are several sources of noise pickup which become evident in DC measurements. In the absence of intrinsic sample noise, it is possible to improve the signal quality further by using special low-noise electronics. The lack of reproducibility of the temperature gradient between different samples (Fig. \ref{dT}) is at least partly due to errors in thermometry attributable to high intrinsic noise and low reproducibility of the $InO _x$ thermometers. This can also be improved using $RuO _2$ or $NbN$ as thermometric material. 

This one-heater-two-thermometer setup is particularly suited for the measurement of 2D superconducting thin films, for which Nernst effect measurement is an important tool to study quantum fluctuations.  The technique can be extended to study Nernst effect in VdW stacks and (by changing to a crystalline substrate like STO) to crystalline materials like cuprate thin films and iron chalcogenides. The low noise level makes it feasible to employ techniques of noise measurement to study fluctuations in thermoelectric effects, which has been predicted to be a useful tool in several branches of in condensed matter physics, including the search for Majorana states in condensed matter systems \cite{Smirnov2018,Smirnov2019}, study of vortex phase transitions in superconductors \cite{Chung2012} and the study of spin-Seebeck effects \cite{Matsuo2018}. 

\vspace{2cm}

\begin{acknowledgments}
The authors would like to thank Pratap Raychaudhuri, Kamran Behnia and Olivier Bourgeois for valuable discussions.  This research was supported by the Israel science foundation (Grant No. 783/17) and NSF-BSF (Grant No. 2017677) 

\end{acknowledgments}

\vspace{2cm}

The data presented in this manuscript are available from the corresponding author upon reasonable request.

\end{document}